\documentclass[pra,twocolumn,floatfix]{revtex4}
\usepackage{amsmath,amssymb,graphicx}
\newcommand{\be}{\begin{equation}}
\newcommand{\ee}{\end{equation}}

\newcommand{\R}{\mathcal{R}}
\newcommand{\T}{\mathcal{T}}
\newcommand{\im}{\,\mathrm{Im}\,}
\newcommand{\re}{\,\mathrm{Re}\,}
\newcommand{\tr}{\,\mathrm{tr}\,}

\begin{document}
\title{A model for chaotic dielectric microresonators}
\author{J. P. Keating}
\author{M. Novaes}
\affiliation{School of Mathematics, University of Bristol, Bristol
BS8 1TW, United Kingdom}
\author{H. Schomerus}
\affiliation{Department of Physics, Lancaster University,
Lancaster LA1 4YB, United Kingdom}

\begin{abstract}
We develop a random-matrix model of two-dimensional dielectric
resonators which combines internal wave chaos with the deterministic
Fresnel laws for reflection and refraction at the interfaces. The
model is used to investigate the statistics of the laser threshold
and line width (lifetime and Petermann factor of the resonances)
when the resonator is filled with an active medium. The laser
threshold decreases for increasing refractive index $n$ and is
smaller for TM polarization than for TE polarization, but is almost
independent of the number of out-coupling modes $N$. The Petermann
factor in the line width of the longest-living resonance also
decreases for increasing $n$ and scales as $\sqrt{N}$, but is less
sensitive to polarization. For resonances of intermediate lifetime,
the Petermann factor scales linearly with $N$. These qualitative
parametric dependencies are consistent with the random-matrix theory
of resonators with small openings. However, for a small refractive
index where the resonators are very open, the details of the
statistics become non-universal. This is demonstrated by comparison
with a particular dynamical model.

\end{abstract}
\maketitle

\section{Introduction}

Two-dimensional dielectric microresonators attract considerable
attention because of their prospective application as microlasers
and single-photon cavities, as well as sensors in chemical and
biological systems
\cite{review1,review2,review3,nockel,Cao,krioukov}. These systems
also allow the study of generic properties of partially confined
waves in a clean setting, and hence provide a route to acquire
knowledge which can be transferred, e.g., to mesoscopic electronic
devices \cite{datta}. The analogy is most complete when comparing
non-interacting electrons in cooled, patterned two-dimensional
semiconductor devices (which usually leak through quantum-point
contacts) with passive optical (or micro-wave) resonators equipped
with small openings. For such geometries, semiclassical methods and
random-matrix theory have provided a deep theoretical understanding
of the spectral and transport properties, which have been found to
agree with numerous experiments and numerical computations
\cite{stoeckmann,efetov,haake,beenakker1997,fyodorovsommers,guhr}.
In particular, it is now well established that the predictions of
random-matrix theory, designed for wave-chaotic systems with strong
mode-mixing, are of universal applicability when the width of the
openings $W$ is much less than the linear system size $L$, implying
a long mean lifetime in the system. For two-dimensional ballistic
geometries, this requirement can be quantified by comparing the
numbers $N\propto W k$ of outcoupled channels at wavenumber $k$ to
the number of channels $M\propto L k$ which are mixed by the
scattering at the boundaries. Standard random-matrix universality
then requires $M\gg N$.

Dielectric microresonators, however, leak everywhere around the
interface and are far more open than the resonators considered in
standard random-matrix theory. The photonic confinement relies on
internal reflection, which only becomes perfect for angles of
incidence above a critical value. As a consequence, the effective
`openness' increases with the system size, resulting in $N\propto
M$. Universality is no longer guaranteed, and a more detailed
modelling is required. This modelling also has to account for the
possible presence of an active medium, essential for microlasers,
which have been manufactured in many forms and materials
\cite{nockel,YS93,gmachl,VKLISLA98,lee1,starofdavid,schwefel,
kneissl,benmessaoud,courvoisier,Lee,Harayama1,Fukushima1,lebental,Tanaka}.

In this paper we develop a quantum-dynamical description of
two-dimensional dielectric microresonators which combines
wave-chaotic propagation of photons inside the resonator with the
Fresnel laws for reflection and refraction at the interface. The
general construction is based on a variant of the quantum
surface-of-section method
\cite{crespi,bogomolny,georgeot,vallejos,almeida,tureci}. The
propagation inside the resonator is expressed in terms of an
internal scattering matrix, which we specify by either using
random-matrix theory \cite{haake,mehta} or a quantum-dynamical
paradigm of chaotic wave propagation, the quantum kicked rotator
\cite{kickedrotator}. Similar models have recently been developed
for mesoscopic electronic and hybrid-superconducting devices
\cite{fyodorov5,jacquod1,tworzydlo,ossipov,ehrenfestreview}, which
require different boundary conditions to open up the system and do
not allow for amplification.

The random-matrix and quantum kicked rotator variants of the model
are explored to investigate the threshold and the quantum-limited
line width of wave-chaotic dielectric microlasers. The laser
threshold is related to the decay rate $\Gamma$ of the longest
living resonances, and can be read off from the imaginary part of
the frequency (which is complex for the non-hermitian operator
describing an open system). For a good, conventional laser resonator
(with almost-perfect mirrors), the line width (full width at half
maximum of the Lorentzian line shape) is given by the Shawlow-Townes
formula \cite{shawlow}
\begin{equation}
\Delta\omega_{ST}=\frac{1}{2}\frac{\Gamma^2}{I}, \label{eq:st}
\end{equation}
where $I$ is the total output intensity. In an open resonator, the
line width is enhanced with respect to this prediction by the
so-called Petermann factor $K$,
\begin{equation}
\Delta\omega=K\,\Delta\omega_{ST}, \label{eq:stk}
\end{equation}
which can be related to the mutual non-orthogonality of the
resonance modes obtained from the non-hermitian operator
\cite{petermann,siegman,patra,frahm,schomerus,berry}, and has been
studied extensively in a wide range of quantum-optical frameworks
\cite{goldberg,brunel,grangier,dutra,bardroff,exter,cheng2,cheng3,hackenbroich,cheng1,eijkelenborg,emile}.
We investigate how $\Gamma$ and $K$ depend on the refractive index
$n$ and the size of the resonator (quantified by $N$ and $M$), and
also discriminate between the two possible polarizations TE and TM
(where the electric or magnetic field lies inside the resonator
plane, respectively). The results are contrasted with predictions of
random-matrix theory for resonators with small openings
\cite{fyodorovsommers,patra,frahm,schomerus,sommers,fyodorov4}.

We start this paper in Section \ref{sec:2} with a brief summary of
the basic  concepts involved in the description of two-dimensional
dielectric microresonators. The quantum-dynamical models are
formulated in Section \ref{sec:3}. In  Section \ref{sec:4} we
present the statistics of decay rates $\Gamma$ (determining the
laser threshold) and Petermann factors $K$ (determining the line
width). These results are obtained by numerical sampling of the
random-matrix ensemble and the parameter space of the quantum-kicked
rotator. A summary of the results and conclusions is presented in
Section \ref{sec:5}.

\section{\label{sec:2} Two-dimensional dielectric resonators}

We consider planar dielectric microresonators of a small width
$\Delta z$ whose material properties are characterized by a
refractive index $n(\omega)$, which may depend on the angular
frequency $\omega$. Amplification is modelled within an effective
medium approach, so that the refractive index is complex, with
$-\im n>0$ proportional to the amplification rate. The refractive
index is taken as homogeneous within the resonator and unity in
the surrounding medium. The geometry of the resonator is specified
by a region $D$ in the $(x,y)$ plane.

We  assume that only the lowest-lying transverse mode is excited,
which provides the best confinement at the planar interfaces and
hence results in the longest-living resonances. For small
wavelengths ($\lambda=c/\omega\ll L$, where $c$ is the velocity of
light and $L$ is the linear dimension of the resonator in the
plane), the radiation outside the resonator is mostly confined to
the resonator plane. Throughout this plane, the electromagnetic
field can then be described by a scalar wavefunction $\psi({\bf
r})$, ${\bf r}=(x,y)$, which represents the electric  or magnetic
field component parallel to the $z$ axis (TM or TE polarization,
respectively).

The wavefunction $\psi(x,y)$ obeys the Helmholtz equation
\begin{equation}
[c^2\nabla^2+n^2({\bf r})\omega^2]\psi({\bf r})=0.\label{eq:helm}
\end{equation}
At the interfaces the wavefunction $\psi$ and its normal
derivative $\partial_\perp\psi$ are continuous for TM
polarization, while for TE polarization these continuity
requirements are fulfilled by $\psi$ and $n^{-2}\partial_\perp
\psi$.

For interfaces with no curvature, these boundary conditions result
in Snell's law
\begin{equation}
\sin\eta=n\sin\chi,\label{eq:snell}
\end{equation} which relates the angle of incidence $\eta$ from
a wave approaching the interface to the angle of refraction $\chi$
of the wave within the refractive medium. The reflection
probabilities for both polarizations are given by the Fresnel laws
\cite{BornWolf,hentschel}
\begin{equation}
R_{TM}=\frac{\sin^2(\chi-\eta)}{\sin^2(\chi+\eta)},\quad
R_{TE}=\frac{\tan^2(\chi-\eta)}{\tan^2(\chi+\eta)}.
\label{eq:fresnel}
\end{equation}
From within the medium, total reflection occurs for  angles of
incidence $\chi>\chi_c$ larger than the critical angle, which is
determined by the condition $\sin\chi_c=1/n$.

For the more complicated interface geometry of a microresonator, the
radiative properties are encoded in the scattering matrix
$S(\omega)$, which relates the incoming and outgoing amplitudes
$\psi^{(\rm out)}_i=\sum_{j}S_{ij}\psi^{(\rm in)}_j$ in a suitably
chosen, flux-normalized basis of scattering states. The poles of the
scattering matrix are defined by the condition $S(\omega_m)=\infty$.
For $\omega=\omega_m$, the Helmholtz equation (\ref{eq:helm})
permits solutions $\psi_m({\bf r})$ with purely outgoing boundary
conditions, $\psi^{(\rm in)}_j=0$.  For a real refractive index,
these poles lie in the lower part of the complex plane $\im
\omega_m\equiv-\Gamma_m/2<0$, where $\Gamma_m$ is the cold-cavity
decay rate of the mode. Amplification shifts the poles towards the
real axis. Within this effective-medium approach, the laser
threshold is reached when the first pole crosses the real axis, so
that the purely outgoing wave field of the laser can be realized at
physical, real frequencies $\omega=\omega_m$
\cite{misirpashaev,zaitsev}.

The active medium within the resonator emits radiation due to the
spontaneous emission of photons and their subsequent amplification.
Assuming total population inversion of the active transition within
the medium, the frequency-resolved output intensity is given by
\cite{kirchhoff1,kirchhoff2}
\begin{equation}
I(\omega)=\frac{1}{2\pi}\tr[S^\dagger(\omega)S(\omega)-1].
\label{eq:i}
\end{equation}
This expression vanishes for a passive medium (with a real
refractive index), for which the scattering matrix is unitary.

Close to the laser threshold, the radiation is dominated by a
Lorentzian peak around $\omega=\re\omega_m$. The width of the peak
can be associated to the spontaneous emission processes, which
perturb the amplitude and phase of the emitted radiation. The
Shawlow-Townes formula (\ref{eq:st}) arises when the resonance in
the intensity (\ref{eq:i})  is calculated within Breit-Wigner
perturbation theory, which is based on the mutually orthogonal modes
of the closed resonator. Equation (\ref{eq:stk}) including the
Petermann factor accounts for the non-orthogonality of the modes in
the open resonator. For $\omega L\gg c$, the Petermann factor can be
expressed in terms of the resonance wave function as
\cite{siegman,remark1},
\begin{equation}
K_m=\frac{\left|\int_{D}d{\bf r}\,|\psi_m({\bf
r})|^2\right|^2}{\left|\int_{D}d{\bf r}\,\psi_m({\bf
r})^2\right|^2}. \label{eq:p0}
\end{equation}

Above the laser threshold, the resonator becomes unstable, since the
gain in the medium outweighs the losses through the interfaces. The
steady-state intensity is limited by pumping and saturation of the
medium, which requires study of the non-linear regime as done, e.g.,
in Refs.\ \cite{dutra,Harayamasunada,hackenbroich}. The feedback
with the medium stabilizes the amplitude of the laser, but not the
phase, whose dynamics gives rise to a finite laser line width. The
resulting width is half the width obtained from the cold-cavity
calculation, which also includes the amplitude fluctuations
\cite{goldberg}. Equations (\ref{eq:st}) and (\ref{eq:stk}) include
this reduction factor of one-half.

\section{\label{sec:3} Quantum-dynamical model}

\subsection{Construction}

In order to set up the quantum-dynamical model of a dielectric
microresonator we separate the motion within the cavity, which is
assumed to be wave-chaotic, from the reflection and refraction
processes at the resonator interfaces, which is assumed to be
governed by the Snell and Fresnel laws (\ref{eq:snell}) and
(\ref{eq:fresnel}), respectively.

The construction follows a variant of the quantum surface-of-section
method \cite{crespi,bogomolny,georgeot,vallejos,almeida} which we
adapt to the specifics of wave propagation in a microresonator, and
bares similarities to the methods used in an efficient numerical
scheme for specific cavities \cite{tureci}. For motivation let us
first consider the classical ray dynamics in  the system. The
successive internal reflections of these rays at the interfaces are
conveniently represented in terms of Birkhoff's canonical
coordinates, given by the length along the boundary, $q$, and the
sine of the  angle of incidence, $p=\sin\chi$. The internal ray
dynamics is then reduced to a sequence of points $(q_n,p_n)$, which
are generated by an area-preserving map ${\cal
M}:(q_n,p_n)\to(q_{n+1},p_{n+1})$. At each encounter with the
interface, a ray is split into a refracted part, which escapes to
the exterior, and a reflected part which remains inside the
resonator. The relative amount of reflection is measured by the
reflection coefficients (\ref{eq:fresnel}), which in Birkhoff
coordinates are written
\begin{subequations}\label{eq:rtmte}
\begin{eqnarray}
R_{TM}(p)&=&\left(\frac{\sqrt{1-p^2}-\sqrt{n^{-2}-p^2}}{\sqrt{1-p^2}+
\sqrt{n^{-2}-p^2}}\right)^2,
\\
R_{TE}(p)&=&\left(\frac{\sqrt{1-p^2}-n\sqrt{1-n^2p^2}}{\sqrt{1-p^2}+
n\sqrt{1-n^2p^2}}\right)^2
\end{eqnarray}
for $|p|<1/n$, while $R=1$ for $|p|>1/n$. \end{subequations}

In wave optics, a similar separation of the internal dynamics and
the encounters with the interface can be carried out for $kL\gg 1$.
In this limit, the internal wave function $\psi^{(\rm
int)}=\psi^{(\rm int,in)}+\psi^{(\rm int,out)}$ can be separated
into the component $\psi^{(\rm int,in)}$ which propagates away from
the interfaces, towards the interior of the resonator, and the
component $\psi^{(\rm int,out)}$ which propagates towards the
interfaces. Because the separation relies on the propagation
direction it is best carried out in momentum space, corresponding to
states with a well defined canonical coordinate $p$. The geometry of
the collision with the interface is encoded in the internal
scattering matrix $F(\omega)$, which relates the components of the
internal wave function,
\begin{equation}
\psi^{(\rm int,out)}=F(\omega)\psi^{(\rm int,in)}. \label{eq:f}
\end{equation}
This is equivalent to Bogomolny's transfer operator
\cite{bogomolny}, originally introduced for closed systems. The
matrix $F(\omega)$ is unitary and symmetric, the latter property
arising from the time-reversal symmetry in a dielectric medium. The
dimension $M\sim {\rm Int}[Cn\omega/\pi c]$ of $F(\omega)$ depends
on the perimeter $C$ of the interface, which is proportional to the
system size, $C\propto L$.

In the classical limit $\omega L/c\propto M\to \infty$,
$F(\omega)$ corresponds to the map ${\cal M}$. When $\omega
L/c\gg 1$, the momentum $p$ is quasi-continuous,
\begin{equation}
p_l\sim \frac{2l-M-1}{2M},\quad l=1,\ldots,M .\label{eq:pl}
\end{equation}

At the interfaces the internal wave function is coupled to the
external wave field, which can be decomposed in the conventional
scattering states $\psi^{\rm (ext)}=\psi^{\rm (ext,in)}+\psi^{\rm
(ext,out)}$.  The coupling is of the general form
\begin{subequations} \label{eq:inout}
\begin{eqnarray}
\psi^{(\rm int,in)}&=&\cal R\psi^{(\rm int,out)}+\cal T\psi^{(\rm
ext,in)},
\\
\psi^{(\rm ext,out)}&=&-\cal R\psi^{(\rm ext,in)}+\cal T\psi^{(\rm
int,out)},
\end{eqnarray}
\end{subequations}
where ${\cal R}$ is the reflection matrix  and $\T$ is the
transmission matrix, constrained by ${\cal R}^\dagger{\cal R}+{\cal
T}^\dagger{\cal T}=1$. The linear Eqs.\ (\ref{eq:f}) and
(\ref{eq:inout}) can be solved to arrive at the scattering matrix
\begin{equation}
S(\omega)=-\R+\T F(\omega)\frac{1}{1-\R F(\omega)}\T. \label{eq:s}
\end{equation}
This general form of the scattering matrix has also been encountered
in recent investigations of electronic transport and superconducting
hybrid structures with small ballistic openings
\cite{fyodorov5,jacquod1,tworzydlo,ossipov,ehrenfestreview}.

Our construction of the quantum-dynamical model is completed by
concrete specifications of the matrices $\R$ and $F(\omega)$. For
$kL\gg1$, Snell's law dictates that the matrices $\R$ and $\T$ in
Eq.\ (\ref{eq:inout}) become diagonal between states which conserve
$np$, while the Fresnel laws (\ref{eq:rtmte}) deliver the values of
the diagonal elements,
\begin{equation}
\R_{lm}=\delta_{lm}\sqrt {R(p_l)}, \quad\T_{lm}=\delta_{lm}\sqrt
{1-R(p_l)},  \label{eq:r}
\end{equation}
with $p_l$  given by Eq.\ (\ref{eq:pl}).

Concerning the internal dynamics encoded in $F(\omega)$, we now
follow two routes --- we either employ random-matrix theory or a
generic wave-chaotic quantum map, the quantum kicked rotator.

\subsubsection{Random-matrix model}

It is commonly accepted that statistical properties of wave systems
with chaotic ray dynamics are well described by random-matrix
theory. In the present case of a dielectric system,  which preserves
time-reversal symmetry, this approach amounts to assuming that $F$
at fixed, real $\omega$ can be represented by a unitary symmetric
matrix randomly drawn from the Circular Orthogonal Ensemble (COE)
\cite{mehta}. This substitution should best capture the statistical
features of the longest-living resonances, since in a realistic
resonator ray chaos is only established after a few internal
reflections.

In general, the frequency dependence of  $F(\omega)$ is complicated.
In the subsequent analysis we will only require the  local
dependence around the poles of the scattering matrix, and employ as
a simple approximation
\begin{equation} F(\omega)=\exp(i n \omega \bar
L/c)U,\quad U\mbox{ from COE$(M)$, fixed}, \label{eq:fu}
\end{equation}
 where $\bar L$
is the mean propagation distance between two reflections. Sabine's
law of room acoustics delivers the universal expression $\bar L=\pi
A/C$, where $A$ is the area of the resonator and $C$ is its
perimeter. Our approximation entails $(2\pi)^{-1}\im \tr
F^\dagger\frac{dF}{d\omega}=M n\bar L/2c\pi\sim A n\omega/2\pi c$.
This reproduces the correct expression for the mean density of
states, while a more elaborate modelling of the $\omega$ dependence
would also describe fluctuations around this value.

The random-matrix model of two-dimensional wave-chaotic
microresonators is obtained by combining  Eqs.\ (\ref{eq:s}),
(\ref{eq:r}), and (\ref{eq:fu}). In order to simplify the notation
we introduce the quasi-energy
\begin{equation} \theta=n \omega
\bar L/c.
\end{equation}
 This delivers the scattering matrix \begin{equation}
S=-\R+\T U\frac{1}{\exp(-i \theta)-\R U }\T, \label{eq:s2}
\end{equation}
where $U$ is an $M\times M$ matrix from the COE, while the
diagonal matrix $\R$ is given in Eq.\ (\ref{eq:r}).

\subsubsection{Quantum kicked rotator model}

The quantum kicked rotator \cite{kickedrotator} is a quantized
version of the classical standard map on the torus
$(q,p)\in[0,1)^2$, which consists of a combination of torsions $p\to
p+q$ and nonlinear `kicks' $q\to q+f(p)$, where $f(p)=f(p+1)=-dv/dp$
is periodic and can be represented as the derivative of a potential
$v(p)$. In order to break all symmetries  apart from time-reversal
symmetry we employ a quantum kicked rotator corresponding to a
torsion followed by a kick with
\begin{eqnarray}
&&f(p)=K_1\sin(2\pi p)+K_2\sin(4\pi p+\alpha),\\
&&v(p)=\frac{K_1}{2\pi}\cos(2\pi p)+\frac{K_2}{4\pi}\cos(4\pi
p+\alpha),
\end{eqnarray}
where the kicking strengths $K_1$, $K_2$ and the shift $\alpha$
are free parameters. The classical dynamics is known to be chaotic for $K_1$, $K_2\gtrsim 8$.

The propagator, or Floquet matrix, takes the form
\begin{equation}
U_{nm}=\frac{1}{\sqrt{iM}}
e^{\frac{i\pi}{M}(m-n)^2-\frac{iM}{2}\left[v\left(\frac{n}{M}\right)+v\left(\frac{m}{M}\right)\right]}.\label{eq:ukr}
\end{equation}
The quantum kicked rotator model of a wave-chaotic microresonator
follows when this matrix is introduced into the scattering matrix
(\ref{eq:s2}).

\subsection{Resonances and the Petermann factor}

We now explore the general structure of the scattering matrix in the
quantum-dynamical model and derive general relations for the
resonances and the Petermann factor.

The scattering matrix (\ref{eq:s}) diverges when the resonant
denominator $1-\R F(\omega)$ has a vanishing eigenvalue. The matrix
$\R F(\omega)$ describes one round trip of a wave which propagates
through the cavity, and then  is internally reflected at the
interface. This quantization condition is analogous to the
scattering-matrix quantization condition in a closed systems
\cite{uzy,dietz}, which is recovered for $\R=1$.

For the specific form (\ref{eq:s2}) of the scattering matrix, the
quantization condition takes the form
\begin{equation}
\R U\Psi_m=\exp(-i \theta_m)\Psi_m, \label{eq:eval}
\end{equation}
which is of the form of an eigenvalue equation with eigenvalue
$\exp(-i \theta_m)$. The matrix $\R U$ can be interpreted as a
reduced round-trip operator. Since $\R U$ is subunitary, $\R U(\R
U)^\dagger=\R\R^\dagger<1$, all eigenvalues are submodular,
$|\exp(-i \theta_m)|<1$; therefore the quasienergies $\theta_m$ have
a negative imaginary part, $\gamma_m\equiv -2\im \theta_m>0$.

The subunitarity of $\R U$ furthermore implies that the right
eigenvectors $\Psi_m$ defined by Eq. (\ref{eq:eval}) are not
mutually orthogonal. However, they form a bi-orthogonal set with the
left eigenvectors,
\begin{equation}
\langle\Phi_l|\Psi_m\rangle=\delta_{lm},\label{eq:biorth}
\end{equation}
where the latter are defined by the adjoint eigenvalue problem
\begin{equation}
\Phi_m^\dagger\R U=\Phi_m^\dagger\exp(-i \theta_m). \label{eq:aeval}
\end{equation}

In the original variables, the eigenphases $\theta_m$ can be
interpreted in two different ways: (1) They are poles $\theta_m=n
\omega_m \bar L/c$ of the scattering matrix for fixed (possibly
complex) refractive index $n$ but variable, generally complex
$\omega_m$, (2) they deliver the threshold condition $\theta_m=n_m
\omega_m' \bar L/c$ of a given resonant state for real $\omega_m'$
and a refractive index $n_m=n'+in''_m$ for which only the real part
is fixed. For the scattering matrix (\ref{eq:s2}), both problems are
intimately related since only the product $n\omega$ enters the
expressions. The cold-cavity poles $\omega_m^{(\rm cold)}$ are
obtained for a real refractive index $n=n'$. The threshold
amplification is then given by
\begin{equation}
n''_m=n'\im \omega_m^{(\rm cold)}/\re \omega_m^{(\rm cold)}.
\label{eq:thresh}
\end{equation}

For realistic dielectric microresonators, this relation is indeed
well established in the semiclassical limit $L\gg \lambda={\rm
Re}\,\omega/c$ \cite{hentschelnoeckel}. In such systems, it is
moreover reasonable to assume that the width of the amplification
window of the active medium is much less than the center of this
window, $\re \omega_m$. Within the amplification window, the
threshold values $n''_m$ then mainly depend on the imaginary part
$\im \omega_m^{(\rm cold)}\equiv -\Gamma_m/2$, where $\Gamma_m$ is
the  cold-cavity decay rate. The laser threshold is hence determined
by the longest-living resonances of the cold cavity, which are
characterized by a small value of $\Gamma_m$.

We now turn to the radiation emitted close to resonance, brought
about by steering the amplification close to the threshold value
(\ref{eq:thresh}), $n''=n''_m+\delta n$, while keeping $\omega=\re
\omega_m^{(\rm cold)}+\delta \omega$ real. We combine both
deviations into the quasienergy deviation
$\theta\approx\theta_m+\delta \theta$. The scattering matrix
(\ref{eq:s2}) can then be evaluated by only keeping the resonant
term in the denominator,
\begin{equation}
S\approx \T
U\Psi_{m}\frac{e^{i\theta_m}}{e^{-i\delta\theta}-1}\Phi_{m}^\dagger\T
\end{equation}

The frequency-resolved output intensity Eq.\ (\ref{eq:i}) takes
the form
\begin{equation}
I(\theta)\approx\frac{1}{2\pi
c}K\frac{4\sinh^2(\im\theta_m)}{|\exp(i\re\delta\theta)-\exp(\im\delta\theta)|^2},
\end{equation}
where
\begin{equation}
K=(\Psi_{m}^\dagger\Psi_{m})(\Phi_{m}^\dagger\Phi_{m}).
\label{eq:p1}
\end{equation}
We now can linearize in $\delta\theta$, which produces the
Lorentzian lineshape
\begin{equation}
I(\theta)\approx\frac{1}{2\pi}K\frac{4\sinh^2(\im\theta_m)}{\re\delta\theta^2+\im\delta\theta^2}.
\end{equation}
Since we are mainly concerned with long living resonances in a large
cavity, $-\im \omega_m=\Gamma_m/2\ll\re\omega$, we further linearize
the numerator and obtain in the frequency domain
\begin{equation}
I(\omega)\approx\frac{1}{2\pi}K\frac{\Gamma_m^2}{(\omega-\re\omega_m^{(\rm
cold)})^2+\Delta\omega^2/4},
\end{equation}
where the full width at half maximum is $\Delta\omega=2 \delta n
\re\omega_m^{(\rm cold)}/n'$. The total intensity is given by
\begin{equation}
I=\int d\omega I(\omega)= K\frac{\Gamma_m^2}{\Delta\omega}.
\end{equation}
This recovers relation  (\ref{eq:stk}) [excluding the factor of
$1/2$ due to the suppression of amplitude fluctuations by the
nonlinear feedback in the lasing regime]. Consequently, $K$ defined
in Eq.\ (\ref{eq:p1}) has to be identified with the Petermann
factor.

Expression (\ref{eq:p1}) assumes that the left and right
eigenvectors are normalized
according to the bi-orthogonality condition (\ref{eq:biorth}).
When this condition is dropped, the Petermann factor takes the
more general form
\begin{equation}
K=\frac{(\Psi_{m}^\dagger\Psi_{m})(\Phi_{m}^\dagger\Phi_{m})}{|\Phi_{m}^\dagger\Psi_{m}|^2}.
\label{eq:pgen1}
\end{equation}

In the present, time-reversal-symmetric situation where $U=U^T$, the
right and left eigenvectors defined by Eqs.\ (\ref{eq:eval}) and
(\ref{eq:aeval}) can be chosen (via a suitable normalization) such
that $\Phi=U^*\Psi^*$. This relation can be further exploited by
decomposing $U=VV^T$ in terms of a unitary matrix $V$, which is
fixed up to transformations $V\to VO$ with an arbitrary orthogonal
matrix $O$. This decomposition allows us to pass to the symmetrized
eigenvalue problem
\begin{subequations} \label{eq:symeval}
\begin{eqnarray}
&&V^T\R V\tilde \Psi_m=\exp(-i \theta_m)\tilde\Psi_m,
\\&&
\tilde\Phi^\dagger_m V^T\R V =\exp(-i
\theta_m)\tilde\Phi_m^\dagger,
\end{eqnarray}
\end{subequations} which is solved by the right and left
eigenvectors $\tilde \Psi=V^T\Psi$ and $\tilde
\Phi=V^T\Phi=\tilde\Psi^*$, respectively. In terms of the
symmetrized eigenvectors, the Petermann factor (\ref{eq:pgen1})
takes the form
\begin{equation}
K=\frac{|\tilde\Psi_{m}^\dagger\tilde\Psi_{m}|^2}{|\tilde\Psi_{m}^T\tilde\Psi_{m}|^2}
. \label{eq:p2}
\end{equation}
This expression is formally analogous to expression (\ref{eq:p0})
for the Petermann factor in terms of the resonance wave function
inside the cavity, but involves the eigenvectors $\tilde\Psi_m$ of
the symmetrized reduced round-trip operator $V^T\R V$. Furthermore,
Eq.\ (\ref{eq:p2}) is formally equivalent to the expression used in
earlier random-matrix theories for cavities with small openings
\cite{patra,frahm,schomerus,fyodorov4,janik,chalker,mehlig1,mehlig2}.

\section{\label{sec:4}Numerical results}

We now employ the quantum-dynamical model with scattering matrix
(\ref{eq:s2}) in order to investigate the statistical properties of
wave-chaotic dielectric microlasers. For fixed values of the
refractive index $n$ and polarization (TE or TM), the matrix $U$ is
either chosen as a random representative from the COE or as the
quantum kicked rotator (\ref{eq:ukr}). The laser threshold follows
from Eq.\ (\ref{eq:thresh}), while the Petermann factor follows from
Eq.\ (\ref{eq:p2}). In order to present the results for the life
times and laser threshold we use the scaled imaginary part
$\gamma_m=-2 \im \theta_m =\Gamma_m \bar L/c$.

In the following we contrast the case of a relatively open resonator
with refractive index $n=1.5$ (close to the value of glass) to a
relatively closed resonator with $n=3.6$ (close to the value of
Ga(Al)As). In the random-matrix model, the computations are based on
$10^5$ realizations of $U$ from the COE with matrix dimension
$M=100$ or $M=200$ \cite{mezzadri}. In the kicked-rotator model, the
same number of matrices with identical dimensions are obtained by
varying the parameters $\alpha$, $K_1$ and $K_2$. We also present
$n$-dependent averages, which are based on $10^4$ realizations of
$U$ in  each of the models.

\begin{figure}[t]
\includegraphics[width=\columnwidth]{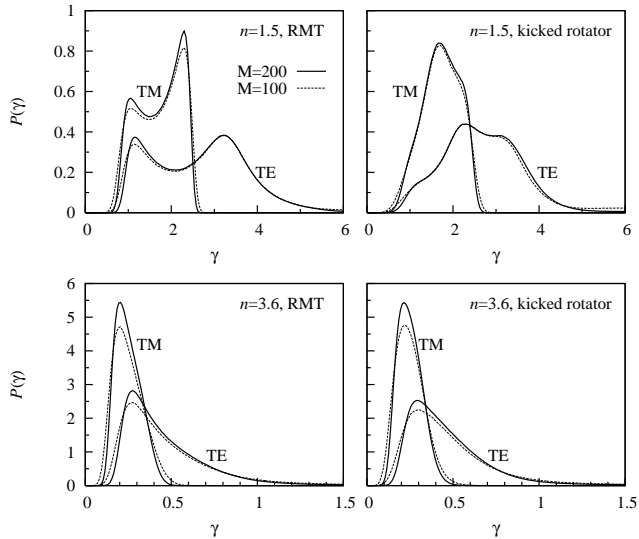}
\caption{Probability distribution $P(\gamma)$ of the scaled decay
rate $\gamma_m=-2\im \theta_m$, obtained from the eigenvalue
equation (\ref{eq:eval}). In the upper panels the refractive index
is $n=1.5$ while in the lower panels $n=3.6$. The left panels are
computed in random-matrix theory, while the right panels are
computed in the quantum kicked rotator model.  The matrix
dimension is $M=200$ (solid curves) and $M=100$ (dashed curves).
The labels TM and TE discriminate the different polarizations.
Each curve is based on $10^5$ realizations of $U$. \label{fig1}}
\end{figure}

Figure \ref{fig1} shows the probability distribution function
$P(\gamma)$ of scaled decay rates for both polarizations and the two
representative values of the refraction index.  The results obtained
for matrices of dimension $M=100$ and 200 are very close, suggesting
a mild dependence of $P(\gamma)$ on the wavelength when the
resonator size increases. For the relatively closed resonator,
$n=3.6$, the distributions found in both models are almost
identical. For the relatively  open resonator with $n=1.5$, however,
the limiting distribution in the random-matrix model is distinctly
different from the result in the quantum kicked rotator model. This
demonstrates that universality starts to break down as the system
becomes more open, due to the greater influence of short-time
dynamics (this breakdown has been studied in more detail for the
spectrum and resonance wave functions of ballistic systems in
\cite{ehrenfestreview,Weyllaw, ourpaper}, using semiclassical
arguments). On the other hand, the characteristic qualitative
features of the distributions display a robust parameter dependence.
For both values of the refractive index $n$ and in both models, the
TE polarization leads to larger decay rates compared to TM. For a
given polarization, the decay rates decrease with increasing $n$.
Both trends are consistent with the general features of the Fresnel
laws (\ref{eq:fresnel}), which provide better confinement for TM
polarization, and for large refractive indices.

\begin{figure}[t]
\includegraphics[width=\columnwidth]{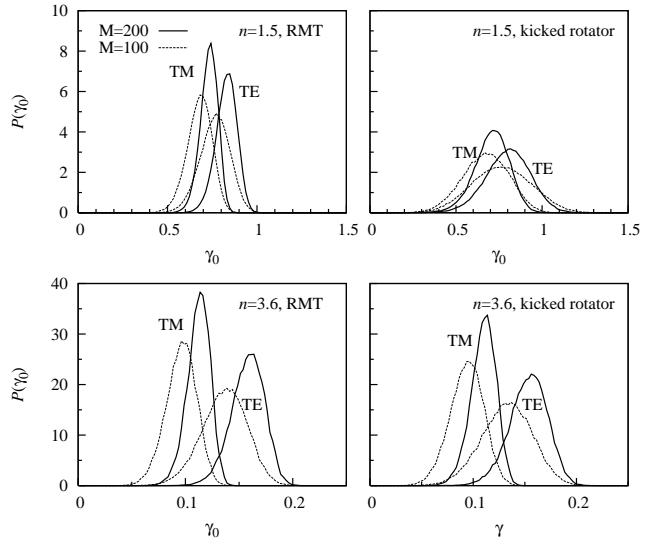}
\caption{Probability distribution $P(\gamma_0)$ of the extremal
(smallest) scaled decay rate $\gamma_0$. The symbols
 and parameters are the same as in Fig.\ \ref{fig1}.
\label{fig2}}
\end{figure}

For each realization, the laser threshold is determined by the
extremal resonance with the smallest rescaled decay rate, which we
denote by $\gamma_0$. Figure \ref{fig2} shows the distribution
$P(\gamma_0)$ of the extremal decay rate for the same parameters
as used in Fig.\ \ref{fig1}. Again the results from both models
coincide for the relatively closed resonator, while for the
relatively open resonator the random-matrix model has a
distinctively more narrow distribution than the quantum kicked
rotator model. In all cases, TM polarization yields smaller decay
rates than TE polarization, which is inherited from the behavior
of $P(\gamma)$ in Fig.\ \ref{fig1}. The results also show a  trend
to larger decay rates as the dimension $M$ is increased. This
trend can be explained by the (hardly visible) sharpening of the
small-$\gamma$ flank in the distribution function $P(\gamma)$
[Fig.\ \ref{fig1}] when $M$ is increased, which suppresses the
tail with very small decay rates. Such a sharpening has also been
demonstrated in the random-matrix theory of cavities with small
ballistic openings, where the distribution function eventually
becomes discontinuous
\cite{fyodorovsommers,sommers,haake2,lehmann}.

\begin{figure}[t]
\includegraphics[width=\columnwidth]{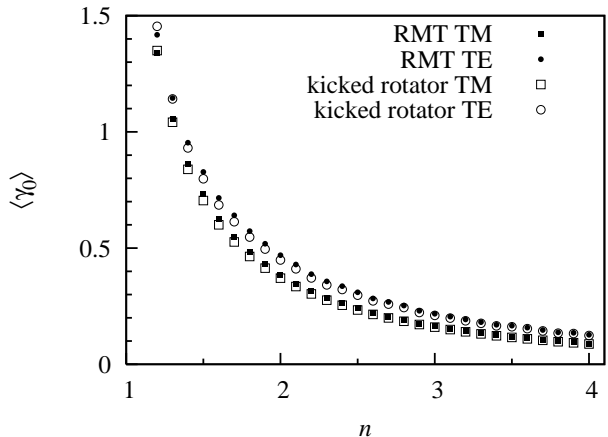}
\caption{Ensemble average $\langle \gamma_0\rangle$ of the scaled
 extremal decay rate $\gamma_0$ as a function of refractive index
$n$, for TM and TE polarization (squares and circles,
respectively). Solid symbols are obtained in random-matrix theory,
while open symbols are obtained in the quantum kicked rotator
model. Each data point is based on $10^4$ realizations of $U$ with
matrix dimension $M=200$. \label{fig3}}
\end{figure}

Figure \ref{fig3} shows the average value $\langle
\gamma_0\rangle$ of the extremal decay rate as a function of the
refractive index $n$. For this quantity, the results from the
random-matrix and kicked-rotator models agree very well. As
expected from the increasing confinement, $\langle
\gamma_0\rangle$ decreases with increasing refractive index $n$,
and is smaller for TM polarization than for TE polarization.

We now turn to the statistics of the Petermann factor. Following the
results from other variants of random-matrix theory
\cite{patra,frahm,schomerus,fyodorov4,janik,chalker,mehlig1,mehlig2}
it has to be expected that the Petermann factor increases with
increasing decay rate of a resonance, and moreover increases with
increasing number of outcoupling channels. In the present class of
systems this number is simply proportional to $M$.

\begin{figure}[t]
\includegraphics[width=\columnwidth]{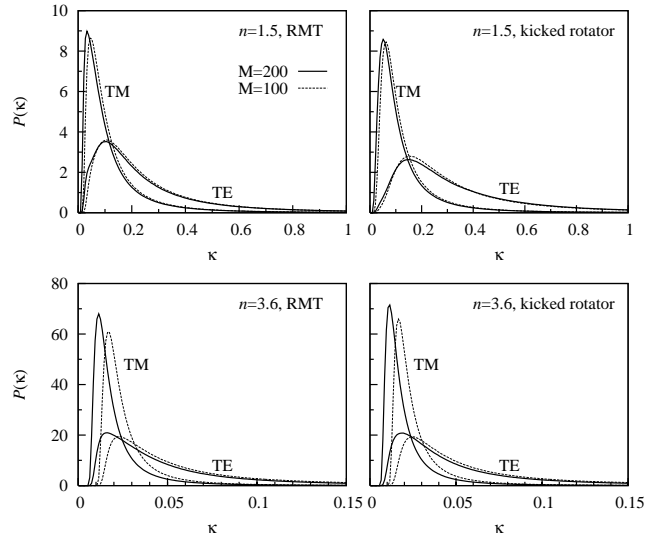}
\caption{Probability distribution $P(\kappa)$ of the scaled
Petermann factor $\kappa=K/M$.  The symbols
 and parameters are the same as in Fig.\ \ref{fig1}. \label{fig4}}
\end{figure}

Anticipating these trends, we show in Fig.\ \ref{fig4} the
distribution $P(\kappa)$ of rescaled Petermann factors $\kappa=K/M$.
The distribution represents the Petermann factor of all resonances,
without discriminating them by width; parameters are the same as in
Fig.\ \ref{fig1}. The collapse of the curves demonstrates clearly
the linear scaling with $M$. Smaller Petermann factors are observed
for TM polarization and/or increasing refractive index, which is
consistent with the reduction of decay rates [see Fig.\ \ref{fig1}].
In contrast to the distribution of decay rates, the results from the
random-matrix and kicked-rotator model agree well for both
refractive indices, indicating that the Petermann factor has a more
universal statistics.

\begin{figure}[t]
\includegraphics[width=\columnwidth]{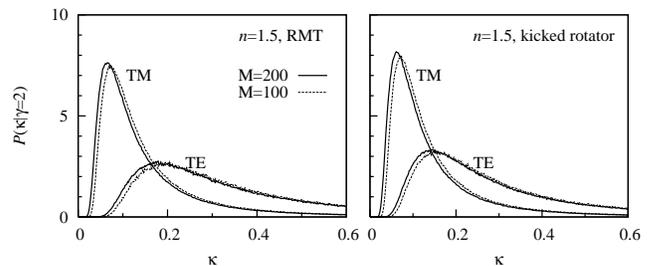}
\caption{Conditional probability distribution $P(\kappa|\gamma)$ of
scaled Petermann factors $\kappa=K/M$ at a fixed value
$\gamma\approx 2.0$ of the scaled decay rate. Results are shown for
both polarizations and two matrix dimensions. The refractive index
is $n=1.5$. \label{fig5}}
\end{figure}

Figure \ref{fig5} shows the conditional probability distribution
$P(\kappa|\gamma)$, of Petermann factors with a given value of the
rescaled decay rate. We again find an approximate linear scaling
with $M$. The dependence on polarization and refractive index
follows the same trends as in Fig.\ \ref{fig4}. Universality is
also present, since the results for both models are very similar.

\begin{figure}[t]
\includegraphics[width=\columnwidth]{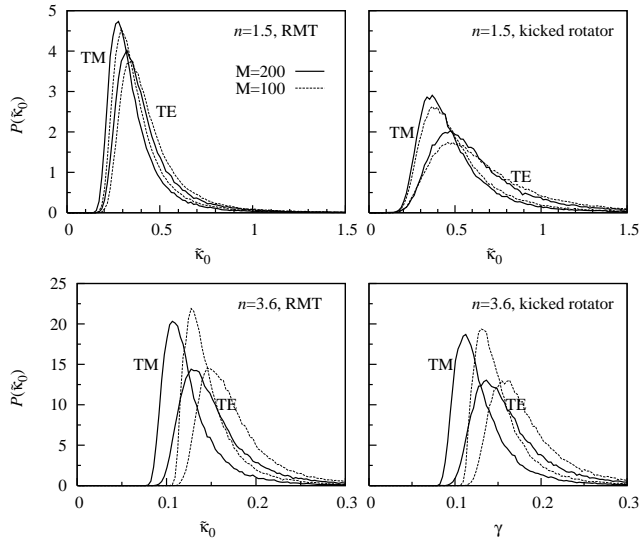}
\caption{(color online) Probability distribution $P(\tilde
\kappa_0)$ of the scaled extremal  Petermann factor $\tilde
\kappa_0=K_0/\sqrt{M}$ of the longest-living resonance.
 The symbols
 and parameters are the same as in Fig.\ \ref{fig1}. \label{fig6}}
\end{figure}

The line width of the microlaser is determined by the Petermann
factor of the longest-living resonance, which we denote by $K_0$
and refer to as the extremal Petermann factor. (For a given
realization, this is not necessarily the smallest Petermann factor
among all the resonances.) As shown in Fig.\ \ref{fig6}, $K_0$
does not scale linearly with $M$, but scales instead with
$\sqrt{M}$. This different parametric dependence can be traced to
the fact that the decay rate of the extremal resonance is not
fixed [Fig.\ \ref{fig2}]. The same parametric dependence, but as a
function of $N$, is also obtained in the random-matrix theory of
cavities with small ballistic openings, where $N\ll M$
\cite{frahm,schomerus}. According to Fig.\ \ref{fig6}, the details
of the statistics for the extremal Petermann factor are again
non-universal for the relatively open resonator with $n=1.5$ ---
the distribution in the random-matrix model is  more narrowly
peaked than in the kicked-rotator model.

\begin{figure}[t]
\includegraphics[width=\columnwidth]{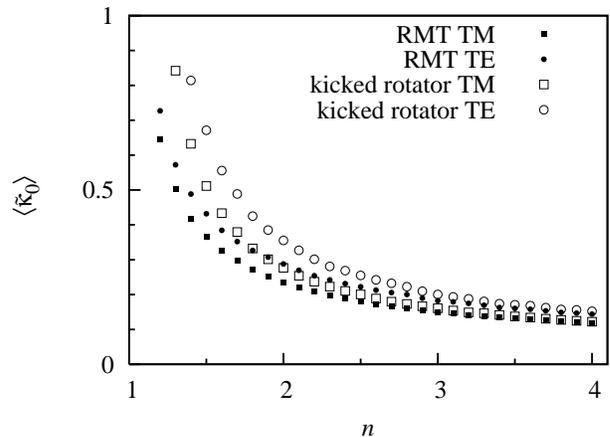}
\caption{Ensemble average $\langle\tilde \kappa_0\rangle$ of the
scaled extremal  Petermann factor $\tilde \kappa_0=K_0/\sqrt{M}$ as
a function of refractive index $n$. The symbols
 and parameters are the same as in Fig.\ \ref{fig3}. \label{fig7}}
\end{figure}

Similarly to what is observed in Fig.\ \ref{fig4}, the extremal
Petermann factor in Fig.\ \ref{fig6} is reduced when increasing
refractive index $n$. However, $K_0$ displays a weaker polarization
dependence. These trends are further underlined in Fig.\ \ref{fig7},
which shows the ensemble average $\langle \tilde \kappa_0\rangle$ of
the scaled extremal Petermann factor $\tilde \kappa_0=
K_0/\sqrt{M}$. This average is significantly larger than the modal
value, which is due to the long tail in the distribution function.
The average decreases with increasing refractive index, similarly to
the averaged extremal decay rate in Fig.\ \ref{fig3}. The results
from the random-matrix and kicked-rotator models converge for large
refractive index.

In summary, Figs.\ \ref{fig1}--\ref{fig3} demonstrate that the laser
threshold of a wave-chaotic dielectric microresonator decreases for
increasing refractive index $n$ and is smaller for TM polarization
than for TE polarization. The threshold increases slowly for
increasing number of out-coupling modes. Figures \ref{fig6} and
\ref{fig7} show that the Petermann factor of a wave-chaotic
dielectric microresonator decreases for increasing refractive index
$n$ and scales as the square-root of the number of out-coupling
modes, but is less sensitive to polarization. Figure \ref{fig4}
shows that this behavior critically depends on the requirement to
determine the longest-living resonance, which first reaches the
laser threshold.

For all quantities, details of the statistics are model-dependent
for relatively open resonators, but become universal for relatively
closed resonators with a large refractive index (for very closed
resonators the results eventually converge to the predictions of
conventional random-matrix theory
\cite{fyodorovsommers,patra,frahm,schomerus,sommers,fyodorov4}).

\section{\label{sec:5}Conclusions}

The quantum-dynamical model of dielectric microresonators presented
in this work addresses geometries which facilitate well-established
wave chaos in the semiclassical limit $\lambda\ll L$,  where $L$ is
the typical resonator dimension and $\lambda$ is the wave length. We
developed two variants of the model, one being based on random
matrix theory while the other is based on the quantum-kicked
rotator. In general, we found that relatively closed resonators
(with a large refractive index) display universal statistics while
model-dependence sets in for relatively open resonators (with a
small refractive index). In the latter case the non-universal
features even affect long-living resonances, which classically
correspond to rays which enjoy many internal reflections at the
dielectric interfaces.

The longest-living resonances determine the laser action of the
resonator. We  concentrated most of our efforts on the laser
threshold and the quantum-limited line width, which should be
directly accessible in a suitable experiment. We found that among
the two competing polarizations, TM will usually win the mode
competition for the most stable resonance. The threshold decreases
with increasing refractive index. The Petermann factor in the line
width also decreases with increasing refractive index, and
furthermore scales $\propto \sqrt{L/\lambda}$.

Experimentally, Petermann factors of the laser mode have been
determined for various geometries in Ref.\
\cite{cheng1,eijkelenborg,emile}. Recent experimental progress
\cite{genack} now makes it possible to address the life time and
Petermann factors of individual resonances via direct means. These
methods are not restricted to the longest living resonance, and do
not require an active medium (they, hence, also apply to passive or
absorbing resonators). Such an experiment could serve to validate
the different scaling $\propto L/\lambda$ of Petermann factors for
resonances of a fixed decay rate.

Many microresonators of interest have a geometry which facilitates
ray chaos only in parts of the classical phase space
\cite{nockel,YS93,gmachl,VKLISLA98,lee1,starofdavid,schwefel,kneissl,benmessaoud,
courvoisier,Lee,Harayama1,Fukushima1,lebental,Tanaka}. While
random-matrix theory is not applicable to these systems, the
kicked-rotator model can readily account for such situations when
the kicking strengths are suitably reduced. In general, Eq.\
(\ref{eq:s}) for the scattering matrix provides a vehicle to study a
wide range of situations, including integrable or disordered
systems, by a suitable round-trip operator ${\cal R}U$. Equation
(\ref{eq:thresh}) for the laser threshold and Eq.\ (\ref{eq:p2}) for
the Petermann factor do not depend on the specific choice of the
round-trip operator and can serve as a starting point for further
analytical considerations.

This work was supported by the EPSRC and by the European
Commission, Marie Curie Excellence Grant MEXT-CT-2005-023778.


\begin{thebibliography}{99}
\bibitem{remark1}  Equation (\ref{eq:p0}) has been originally derived
for paraxial resonators, where the wave function refers to the
transverse profile of the modes \cite{petermann,siegman}, but also
applies to two-dimensional dielectric resonators for which it can be
derived directly from the Helmholtz equation.
\end{thebibliography}
\end{document}